\documentclass[%
reprint,
superscriptaddress,
showpacs,preprintnumbers,
 amsmath,amssymb,
 aps,
longbibliography,
pra,
floatfix,
]{revtex4-2}
\usepackage{float}
\usepackage{xspace}
\usepackage[utf8]{inputenc}
\usepackage[final]{graphicx}
\usepackage{dcolumn}
\usepackage{bm}
\usepackage{color}
\usepackage{verbatim}
\usepackage{soul}
\usepackage{multirow}
\usepackage{hyperref}
 \usepackage{amssymb}
 \usepackage{nicefrac}
\usepackage{upgreek}
\hypersetup{
    unicode=false,          
    pdftoolbar=true,        
    pdfmenubar=true,        
    pdffitwindow=false,     
    pdfstartview={FitH},    
    pdftitle={Phonon anomalies in FeS},    
    pdfnewwindow=true,      
    colorlinks=true,       
    linkcolor=blue,          
    citecolor=blue,        
    filecolor=blue,      
    urlcolor=blue,       
    plainpages=false,        
}

\usepackage{CJK}
\newcommand{\Alg}{\texorpdfstring{\ensuremath{A_{1g}}\xspace}{A1g}}
\newcommand{\Ag}{\texorpdfstring{\ensuremath{A_{g}}\xspace}{Ag}}

\newcommand{\Eg}{\texorpdfstring{\ensuremath{E_{g}}\xspace}{Eg}}

\usepackage{array}
\newcolumntype{L}[1]{>{\raggedright\let\newline\\\arraybackslash\hspace{0pt}}m{#1}}

\newcommand{\vei}{\mbox{{\bf e}$_{\rm i}$}\xspace}
\newcommand{\ves}{\mbox{{\bf e}$_{\rm s}$}\xspace}
\newcommand{\wn}{\ensuremath{\rm cm^{-1}}\xspace}
\DeclareUnicodeCharacter{2212}{-}

\begin{document}

\title{Probing charge density wave phases and the Mott transition in $1T$-TaS$_2$ by inelastic light scattering}
\date{\today}
\author{S. Djurdji\'{c} Mijin}
\affiliation{Center for Solid State Physics and New Materials, Institute of Physics Belgrade, University of Belgrade, Pregrevica 118, 11080 Belgrade, Serbia}
\author{A. Baum}
\affiliation{Walther Meissner Institut, Bayerische Akademie der Wissenschaften, 85748 Garching, Germany}
\author{J. Bekaert}
\affiliation{Department of Physics, University of Antwerp, Groenenborgerlaan 171, B-2020 Antwerp, Belgium}
\author{A. \v{S}olaji\'{c}}
\affiliation{Center for Solid State Physics and New Materials, Institute of Physics Belgrade, University of Belgrade, Pregrevica 118, 11080 Belgrade, Serbia}
\author{J. Pe\v{s}i\'{c}}
\affiliation{Center for Solid State Physics and New Materials, Institute of Physics Belgrade, University of Belgrade, Pregrevica 118, 11080 Belgrade, Serbia}
\author{Y.~Liu} 
\altaffiliation{Present address: Los Alamos National Laboratory, Los Alamos, New Mexico 87545, USA}
\affiliation{Condensed Matter Physics and Materials Science Department, Brookhaven National Laboratory, Upton, NY 11973-5000, USA}
\author{Ge He}
\affiliation{Walther Meissner Institut, Bayerische Akademie der Wissenschaften, 85748 Garching, Germany}
\author{M. V. Milo\v{s}evi\'{c}}
\affiliation{Department of Physics, University of Antwerp, Groenenborgerlaan 171, B-2020 Antwerp, Belgium}
\author{C.~Petrovic}
\affiliation{Condensed Matter Physics and Materials Science Department, Brookhaven National Laboratory, Upton, NY 11973-5000, USA}
\author{Z. V.~Popovi\'{c}}
\affiliation{Center for Solid State Physics and New Materials, Institute of Physics Belgrade, University of Belgrade, Pregrevica 118, 11080 Belgrade, Serbia}
\affiliation{Serbian Academy of Sciences and Arts, Knez Mihailova 35, 11000 Belgrade, Serbia}

\author{R. Hackl}
\affiliation{Walther Meissner Institut, Bayerische Akademie der Wissenschaften, 85748 Garching, Germany}
\author{N.~Lazarevi\'{c}}
\affiliation{Center for Solid State Physics and New Materials, Institute of Physics Belgrade, University of Belgrade, Pregrevica 118, 11080 Belgrade, Serbia}

\begin{abstract}
We present a polarization-resolved, high-resolution Raman scattering study of the three consecutive charge density wave (CDW) regimes in $1T$-TaS$_2$ single crystals, supported by \textit{ab initio} calculations. Our analysis of the spectra within the low-temperature commensurate (C-CDW) regime shows $\mathrm{P3}$ symmetry of the system, thus excluding the previously proposed triclinic stacking of the ``star-of-David'' structure, and promoting trigonal or hexagonal stacking instead. The spectra of the high-temperature incommensurate (IC-CDW) phase directly project the phonon density of states due to the breaking of the translational invariance, supplemented by sizeable electron-phonon coupling. Between 200 and 352\,K, our Raman spectra show contributions from both the IC-CDW and the C-CDW phase, indicating their coexistence in the so-called nearly-commensurate (NC-CDW) phase. The temperature-dependence of the symmetry-resolved Raman conductivity indicates the stepwise reduction of the density of states in the CDW phases, followed by a Mott transition within the C-CDW phase. We determine the size of the Mott gap to be $\Omega_{\rm gap}\approx 170-190$ meV, and track its temperature dependence.
\end{abstract}
\pacs{%
}
\maketitle



\section{Introduction}

Quasi-two-dimensional transition metal dichalcogenides (TMDs) have been in the focus of various scientific investigations over the last 30 years, mostly due to the plethora of charge density wave (CDW) phases. Among all TMD compounds $1T$-TaS$_2$ stands out because of its unique and rich electronic phase diagram \cite{Wilson1975_AP24_117,ScrubyPhilMag1975_255,Thompson1994_PRB49_16899,WenChinPhysB_2019_058504}. It experiences phase transitions at relatively high temperatures,  making it easily accessible for investigation and, mainly for the hysteresis effects, attractive for potential applications such as data storage \cite{Svetin_2017_46048}, information processing \cite{Svetin_2014_103201} or voltage-controlled oscillators \cite{Liu_2016_845}.

The cascade of phase transitions as a function of temperature includes the transition from the normal metallic to the incommensurate CDW (IC-CDW) phase, the nearly-commensurate CDW (NC-CDW) phase and the commensurate CDW (C-CDW) phase occurring at around $T_{IC}= 554$\,K, $T_{NC}= 355$\,K and in the temperature range from $T_{C\downarrow}= 180$\,K to $T_{C\uparrow}= 230$\,K, respectively. Recent studies indicate the possibility of yet another phase transition in $1T$-TaS$_2$ at $T_H$=80\,K, named the hidden CDW state \cite{Salgado_2019_037001, Wang2019_ArXive, stojchevska_2014_177}. This discovery led to a new boost in attention for $1T$-TaS$_2$.

Upon lowering the temperature to $T_{IC}= 554$\,K, the normal metallic state structure, described by the space group $\mathrm{P\bar{3}m1}$ ($\mathrm{D^{d}_{3d}}$), \cite{Gasparov:2002} transforms into the IC-CDW state. As will be demonstrated here, the IC-CDW domains shrink upon further temperature reduction until they gradually disappear, giving place to the C-CDW ordered state. This region in the phase diagram between 554\,K and roughly 200\,K is characterized by the coexistence of the IC-CDW and C-CDW phases and is often refereed to as NC-CDW. At the transition temperature $T_C$ IC-CDW domains completely vanish \cite{PhysRevB.93.214109} and a new lattice symmetry is established.
There is general consensus about the formation of ``star-of-David'' clusters with in-plane $\sqrt{13}a$ $\times$ $\sqrt{13}a$ lattice reconstruction, whereby twelve Ta atoms are grouped around the $\mathrm{13^{th}}$ Ta atom \cite{uchida_1981_393, BROUWER198051}. In spite of extensive investigations, both experimental and theoretical, it remains an open question whether the stacking  of ``star-of-David" clusters is triclinic, trigonal, hexagonal or a combination thereof \cite{BROUWER198051,uchida_1981_393,duffay_1976_617,hirata_2001_361,ramos_2019_165414}. The C-CDW phase is believed to be an insulator \cite{Wilson1975_AP24_117,SiposNatMat_2008_960, Fazekas_1979_229, martino2020} with a gap of around 100\,meV \cite{Gasparov:2002}. Very recent theoretical studies based on density-functional theory (DFT) find an additional ordering pattern along the crystallographic $c$-axis which renders the material three-dimensional metallic. The related has a width of approximately 0.5\,eV along $k_z$ and becomes gapped at the Fermi energy $E_{\rm F}$ in the C-CDW phase \cite{Darancet:2014,LeeSH:2019}.

Nearly all of the previously reported results for optical phonons in $1T$-TaS$_2$ are based on Raman spectroscopy on the C-CDW phase and on temperature-dependent measurements in a narrow range around the NC-CDW to C-CDW phase transition \cite{uchida_1981_393, Gasparov:2002, duffay_1976_617, hirata_2001_361, ramos_2019_165414}. In this article we present temperature-dependent polarization-resolved Raman measurements in the temperature range from 4\,K to 370\,K covering all three CDW regimes of $1T$-TaS$_2$.
Our analysis of the C-CDW phase confirms the symmetry to be P3, while the NC-CDW phase is confirmed as a mixed regime of commensurate and incommensurate domains. The spectra of the IC-CDW phase mainly project the phonon density of states due to breaking of translation invariance and sizeable electron-phonon coupling. The growth of the CDW gap upon cooling, followed by the opening of the Mott gap is traced via the initial slope of the symmetry-resolved spectra. The size of 170-190\,meV and the temperature dependence of the Mott gap are directly determined from high-energy Raman data.

\section{Experimental and numerical methods} 

The preparation of the studied $1T$-TaS$_2$ single crystals is described elsewhere \cite{PhysRevB.97.195117, Li_2012, doi:10.1063/1.4805003, PhysRevB.88.115145}. Calibrated customized Raman scattering equipment was used to obtain the spectra. Temperature-dependent measurements were performed with sample attached to the cold finger of a He-flow cryostat. All measurements were performed under a high vacuum of approximately  $5 \cdot 10^{-5}$ Pa. The 575\,nm laser line of a diode-pumped Coherent GENESIS MX-SLM solid state laser was used as an excitation source. Additional measurements with the 458\,nm and 514\,nm laser lines were performed with a Coherent Innova 304C Argon ion laser.
All spectra shown are corrected for the sensitivity of the instrument and the Bose factor, yielding the imaginary part of the Raman susceptibility $R\chi^{\prime\prime}$ where $R$ is an experimental constant. The linear polarizations of the incident and scattered light are denoted as \vei and \ves, respectively, and are always perpendicular to the $c$-axis.
Low energy data up to 550\,\wn were acquired in steps of $\Delta \Omega = 1\,\wn$ with a resolution of $\sigma \approx 3\,\wn$. The symmetric phonon lines were modelled using Voigt profiles where the width of the Gaussian part is given by $\sigma$. For spectra up to higher energies the step width and resolution were set at $\Delta \Omega = 50\,\wn$ and $\sigma \approx 20\,\wn$, respectively.
The Raman tensors for the $\mathrm{D}_\mathrm{3d}$ point group are given in Table~\ref{table1}. Accordingly, parallel linear polarizations project both \Alg and \Eg symmetries, while crossed linear polarizations only project \Eg. The pure \Alg response then can be extracted by subtraction.

We have performed DFT calculations as implemented in the ABINIT package \cite{Gonze20092582}. We have used the Perdew-Burke-Ernzerhof (PBE) functional, an energy cutoff of 50 Ha for the planewave basis, and we have included spin-orbit coupling by means of fully relativistic Goedecker pseudopotentials \cite{PhysRevB.54.1703,Krack2005}, where Ta-5d$^3$6s$^2$ and S-3s$^2$3p$^4$ states are treated as valence electrons. The crystal structure was relaxed so that forces on each atom were below 10 $\upmu$eV/\AA~and the total stress on the unit cell below 1 bar, yielding lattice parameters $a=3.44$ \AA~and $c=6.83$ \AA. Subsequently, the phonons and the electron-phonon coupling (EPC) were obtained from density functional perturbation theory (DFPT) calculations, also within ABINIT \cite{PhysRevLett.68.3603}. Here, we have used an $18 \times 18 \times 12$ $\textbf{k}$-point grid for the electron wave vectors and a $6 \times 6 \times 4$ $\textbf{q}$-point grid for the phonon wave vectors. For the electronic occupation we employed Fermi-Dirac smearing with broadening factor $\sigma_\mathrm{FD}=0.01$ Ha, which is sufficiently high to avoid unstable phonon modes related to the CDW phases.
\begin{table}[t]
\caption{Raman tensors for trigonal systems (point group $\mathrm{D}_\mathrm{3d}$)}
\label{table1}
\resizebox{\linewidth}{!}{
\begin{tabular}{c c}
$
\Alg = \begin{pmatrix}
a&0&0\\
0&a&0\\
0&0&b\\\end{pmatrix}
$
&
${}^1\Eg = \begin{pmatrix}
c&0&0\\
0&-c&d\\
0&d&0\\
\end{pmatrix}$
${}^2\Eg = \begin{pmatrix}
0&-c&-d\\
-c&0&0\\
-d&0&0\\ \end{pmatrix}$

\\[1mm]
\end{tabular}}
\end{table}
\section{Results and Discussion}
\subsection{Lattice dynamics of the charge-density wave regimes}

\subsubsection{C-CDW phase}

\begin{figure}[t]
 \centering
 \includegraphics[width=85mm]{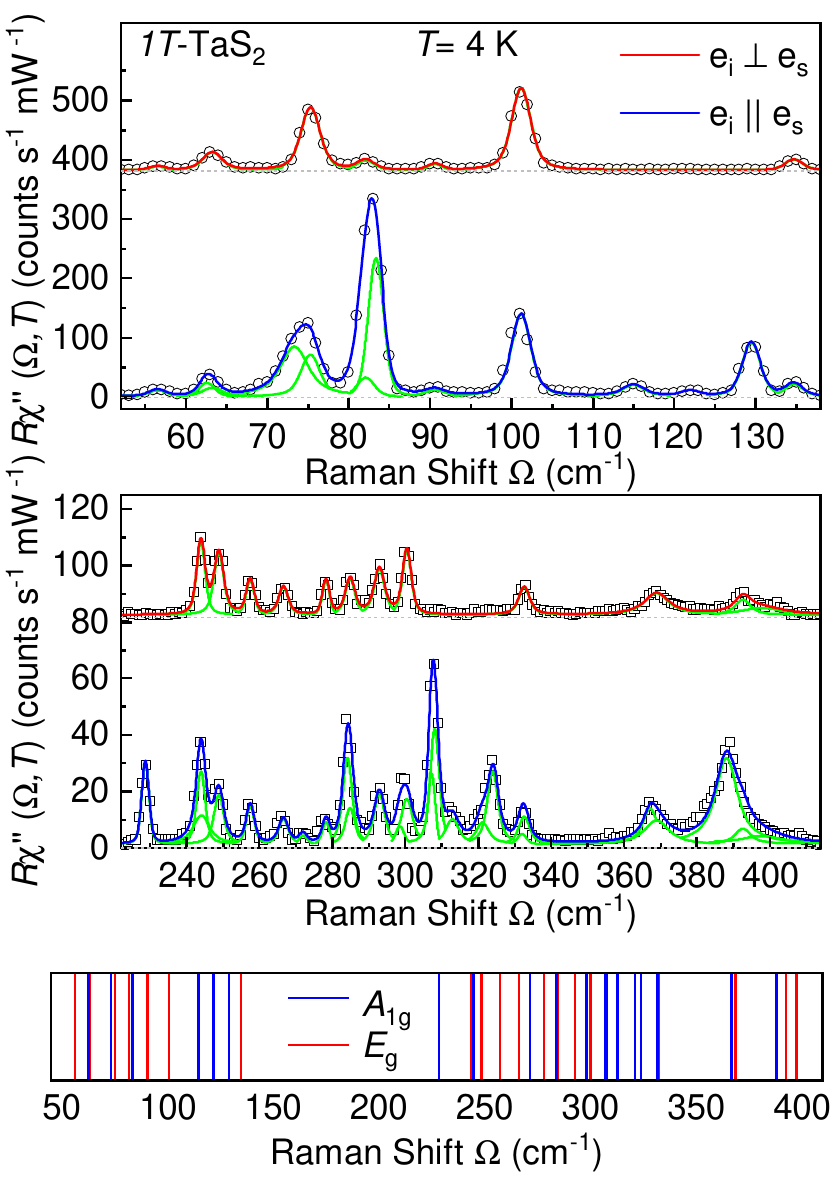}
 \caption{Raman spectra at $T = 4\, \mathrm{K}$, i.e. in the C-CDW phase, for parallel and crossed light polarizations. Red and blue solid lines represent fits of the experimental data using Voigt profiles. Spectra are offset for clarity. The exact energy values are presented in Table~\ref{tableA1}. }
 \label{ref:Figure1}
\end{figure}

\begin{figure}[t]
 \centering
  \includegraphics[width=85mm]{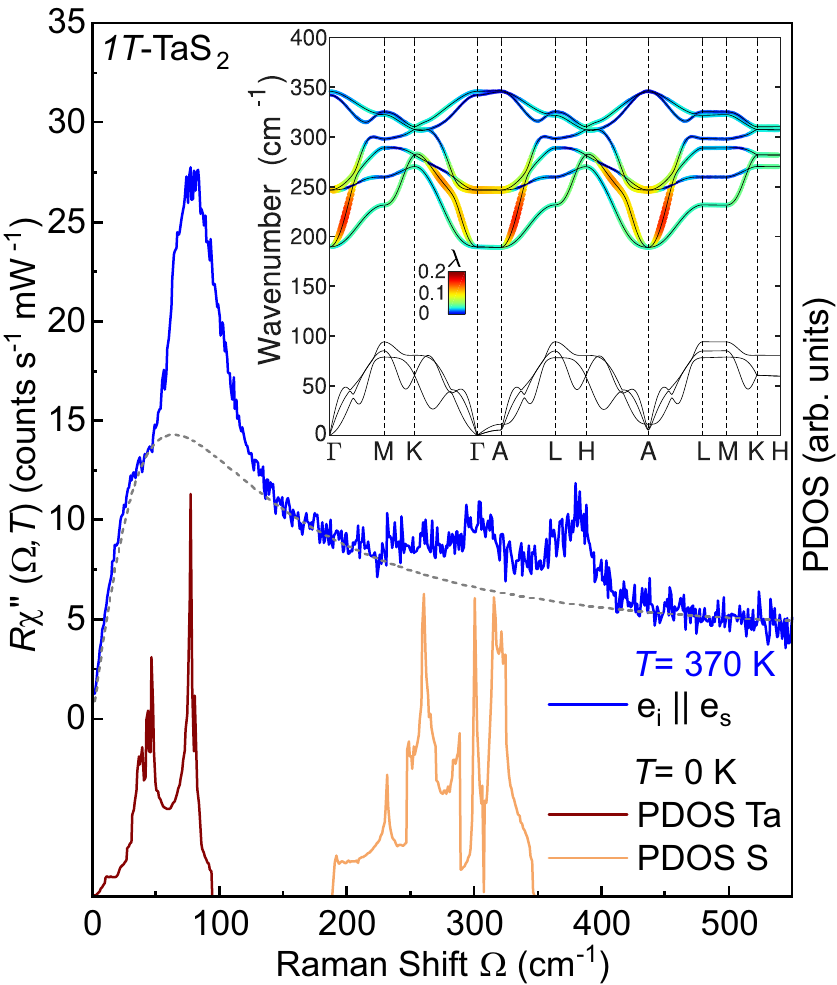}
  \caption{Raman response for parallel light polarizations in the IC-CDW phase at 370 K (blue line). The dashed line depicts the possible electronic continuum. The contributions of the Ta- (dark brown) and S atoms (light brown) to the calculated PDOS are shown below. The inset shows the calculated phonon dispersion of $1T$-TaS$_2$ in the simple metallic phase, with the electron-phonon coupling ($\lambda$) of the optical branches indicated through the color scale.}
 \label{ref:Figure2}
\end{figure}

At the lowest temperatures $1T$-TaS$_2$ exists in the commensurate C-CDW phase. Here, the atoms form so called ``Star-of-David'' clusters. Different studies report either triclinic stacking of these clusters leading to $\mathrm{P\bar1}$ unit cell symmetry \cite{BROUWER198051}, or trigonal or hexagonal stacking and $\mathrm{P3}$ unit cell symmetry \cite{uchida_1981_393, duffay_1976_617, hirata_2001_361, ramos_2019_165414}.
Factor group analysis predicts 57 \Ag \, Raman-active modes with identical polarization-dependence for $\mathrm{P\bar1}$ unit cell symmetry, and alternatively 19 \Alg + 19 \Eg \, Raman-active modes for $\mathrm{P3}$ unit cell symmetry. Our polarized Raman scattering measurements at $T = 4\,\mathrm{K}$, measured in two scattering channels, together with the corresponding cumulative fits are shown in Figure~\ref{ref:Figure1}. As it can be seen, we have observed modes of two different symmetries in the related scattering channels. This result indicates trigonal or hexagonal stacking of the ``Star-of-David'' clusters. The symmetric phonon lines can be described by Voigt profiles, the best fit of which is shown as blue (for parallel light polarizations) and red (crossed polarizations) lines.
After fitting Voigt profiles to the Raman spectra, 38 phonon modes were singled out. Following the selection rules for \Alg \, and \Eg \, symmetry modes, 19 were assigned as \Alg and 19 as \Eg  symmetry, meaning all expected modes could be identified. The contribution from each mode to the cumulative fit is presented in Figure~\ref{ref:Figure1} as green lines, whereas he complete list of the corresponding phonon energies can be found  in the Table~\ref{tableA1} of the Appendix A.

\subsubsection{IC-CDW phase}

At the highest experimentally accessible temperatures $1T$-TaS$_2$ adopts the IC-CDW phase.
Data collected by Raman scattering at \textit{T} = 370\,K, containing all symmetries, is shown as a blue solid line in Figure~\ref{ref:Figure2}. As $1T$-TaS$_2$ is metallic in this phase \cite{PhysRevLett.122.106404} we expect the phonon lines to be superimposed on a continuum of electron-hole excitations which we approximate using a Drude spectrum shown as a dashed line.\cite{Zawadowski:1990,LazarevicHackl2020}

\begin{figure}[t]
\centering
  \includegraphics[width=85mm]{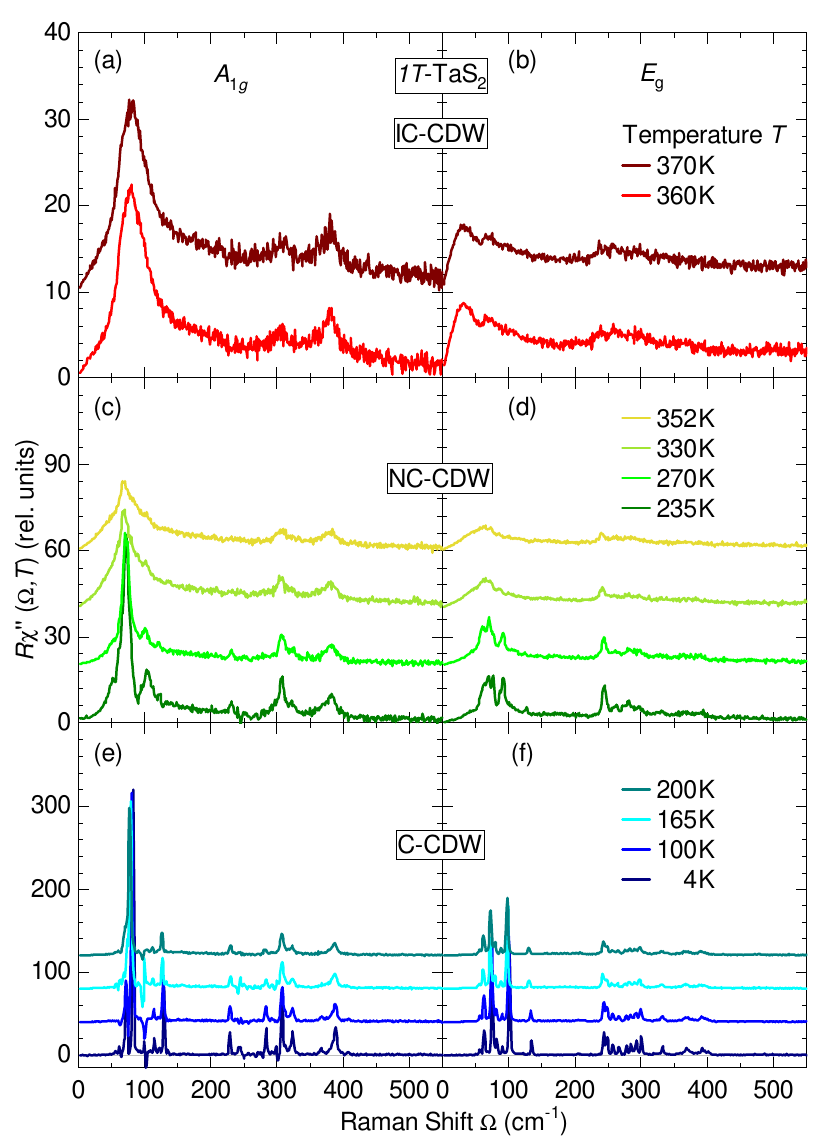}
 \caption{Symmetry-resolved Raman spectra at temperatures as indicated. Both C-CDW (blue lines) and IC-CDW (red lines) domains yield significant contributions to the Raman spectra of the NC-CDW phase (green lines).}
 \label{ref:Figure3}
\end{figure}

Since the IC-CDW phase arises from the normal metallic phase, described by space group $\mathrm{P\bar{3}m1}$,\cite{Rossnagel_2011, Gasparov:2002} it is interesting to compare our Raman results on the IC-CDW phase to an \textit{ab initio} calculation of the phonon dispersion in the normal phase, shown as inset in Fig.~\ref{ref:Figure2}. Four different optical modes were obtained at $\Gamma$: $E_u$ at 189\,cm$^{-1}$ (double-degenerate), $E_g$ at 247\,cm$^{-1}$ (double-degenerate), $A_{2u}$ at 342\,cm$^{-1}$ and $A_{1g}$ at 346\,cm$^{-1}$. Factor group analysis shows that two of these are Raman-active, namely $E_{g}$ and $A_{1g}$ \cite{Gasparov:2002}.

We observe that the calculated phonon eigenvalues of the simple metallic phase at $\Gamma$ do not closely match the observed peaks in the experimental spectra of the IC-CDW phase. Rather, these correspond better to the calculated phonon density of states (PDOS), depicted in Fig.~\ref{ref:Figure2}. As the momentum transfer by light scattering is negligible, the projection of the PDOS requires a way to transfer momentum. We rule out chemical impurity scattering, expected to exist at all temperatures, as the low-temperature spectra (Fig.~\ref{ref:Figure1}) show no signs thereof. The additional scattering channel may come from the electron-phonon coupling (EPC). The calculated EPC, $\lambda$, in the optical modes (inset of Fig.~\ref{ref:Figure2}) is limited, yet not negligible, reaching maxima of $\sim 0.2$ in the lower optical branches around Brillouin zone points $\Gamma$ and A. The Ta-based acoustic modes display several dips that are signatures of the latent CDW phases, for which the EPC cannot be reliably determined. Significant EPC in the optical modes of $1T$-TaS$_2$ is furthermore supported by experimental results linking a sharp increase in the resistivity above the IC-CDW transition temperature to the EPC \cite{Rossnagel_2011}. It also corroborates calculated \cite{PhysRevB.93.214109} and experimentally obtained \cite{Gasparov:2002} values of the CDW gap, which correspond to intermediate to strong EPC \cite{Rossnagel_2011}. Although EPC certainly contributes we believe that the majority of the additional scattering channels can be traced back to the incommensurate breaking of the translational invariance upon entering IC-CDW. Thus the ''weighted'' PDOS is projected into the Raman spectrum (see Fig.~\ref{ref:Figure3} (a) and (b)). These  ''weighting'' factors depend on the specific symmetries along the phonon branches as well as the ''new periodicity'' and go well beyond the scope of this paper.

\subsubsection{NC-CDW phase}

The nearly-commensurate phase is seen as a mixed phase consisting of regions of commensurate and incommensurate CDWs \cite{PhysRevB.56.13757,PhysRevB.94.201108}. This coexistence of high and low-temperature phases is observable in our temperature dependent data as shown in Fig.~\ref{ref:Figure3}. The spectra for the IC-CDW (red curves) and C-CDW phase (blue curves) are distinctly different, as also visible in the data shown above (Figs.~\ref{ref:Figure1} and \ref{ref:Figure2}). The spectra of the NC-CDW phase ($235\,\mathrm{K} < T < 352\,\mathrm{K}$) comprise contributions from both phases. As 352\,K is the highest temperature at which the contributions from the C-CDW phase can be observed in the spectra, we suggest that the phase transition temperature from IC-CDW to NC-CDW phase is somewhere in between 352\,K and 360\,K. This conclusion is in good agreement with experimental results regarding this transition \cite{Thompson1994_PRB49_16899,ScrubyPhilMag1975_255,WenChinPhysB_2019_058504}.



\subsection{Gap evolution}


The opening of a typically momentum-dependent gap in the electronic excitation spectrum is a fundamental property of CDW systems which has also been observed in $1T$-TaS$_2$ \cite{Rossnagel_2011, RevModPhys.60.1129, Gasparov:2002}. Here, in addition to the CDW, a Mott transition at the onset of the C-CDW phase leads to an additional gap opening in the bands close to the the $\Gamma$ point \cite{Sohrt2014, Nat.Mat.Forro}. Symmetry-resolved Raman spectroscopy can provide additional information here using the momentum resolution provided by the selection rules.
\begin{figure}[t]
 \centering
  \includegraphics[width=85mm]{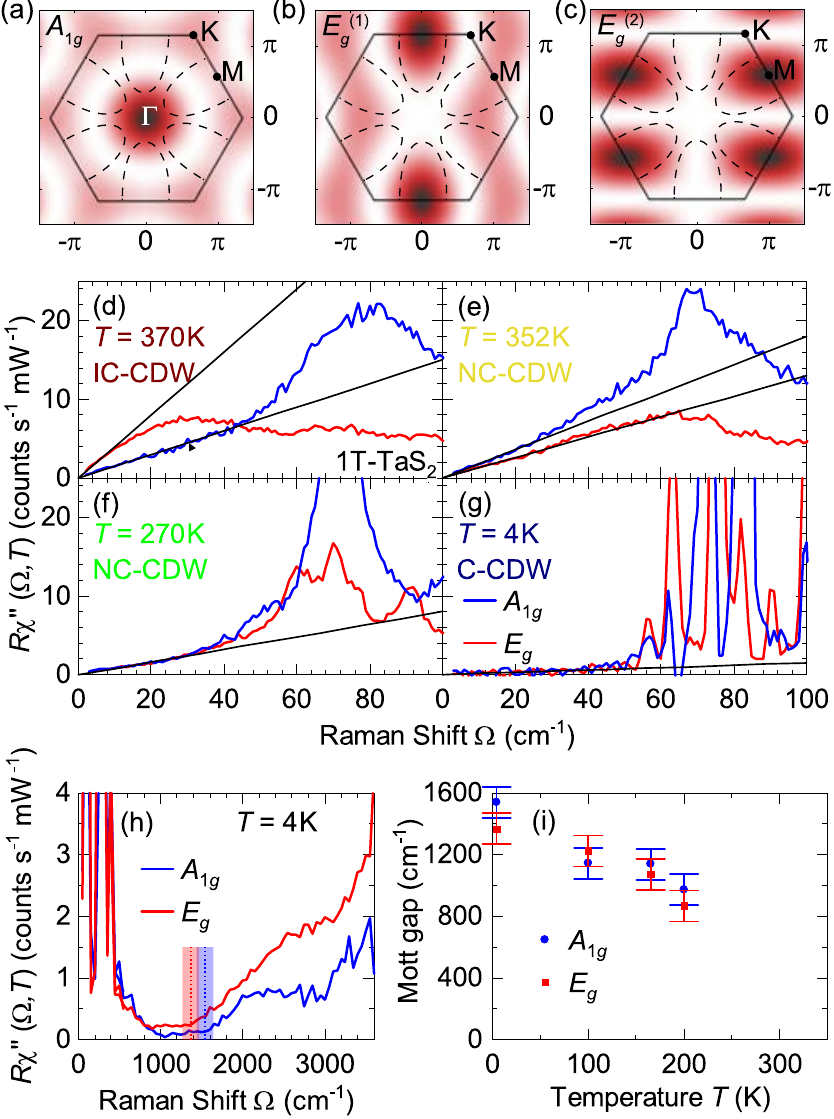}
  \caption{Evolution of the gaps. (a-c) Raman vertices and Fermi surface of $1T$-TaS$_{2}$ for the indicated symmetries. (d-g) Low energy Raman spectra for \Alg symmetry (blue) and \Eg symmetries (red) at temperatures as indicated. The spectra shown are zooms on the data shown in Fig.~\ref{ref:Figure3}. The black lines highlight the initial slope of the spectra. (h) High energy spectra at 4\,K. Vertical dashed lines and colored bars indicate the approximate size and error bar of the Mott gap for the correspondingly colored spectrum. (i) Temperature dependence of the Mott gap $\Delta_\mu$ ($\mu = \Alg, \Eg$)}
 \label{ref:Figure4}
\end{figure}

As shown in Fig.~\ref{ref:Figure4}(a-c), different symmetries project individual parts of the Brillouin zone (BZ). The \Alg vertex mainly highlights the area around the $\Gamma$ point while the \Eg vertices predominantly project the BZ boundaries. The opening of a gap at the Fermi level reduces the density of states $N_{\rm F}$, leading to an increase of the resistivity in the case of $1T$-TaS$_2$. This reduction of $N_{\rm F}$ manifests itself also in the Raman spectra which, to zeroth order, are proportional to $N_{\rm F}$. As a result the initial slope changes as shown Figs.~\ref{ref:Figure4}(d-e), which zooms in on the low energy region of the spectra from Fig.~\ref{ref:Figure3}.
The initial slope of the Raman response is $R\lim_{\Omega\to0}\frac{\partial \chi^{\prime\prime}}{\partial \Omega} \propto N_{\rm F}\tau_0$, where $R$ incorporates all experimental factors. The electronic relaxation $\Gamma^\ast_{0} \propto ({N_{\rm F}\tau_0})^{-1}$ is proportional to the dc resistivity $\rho(T)$. The black lines in Fig.~\ref{ref:Figure4}(d-g) represent the initial slopes and their temperature dependences of the low-energy spectra. The lines comprise carrier relaxation and gap effects, and we focus only on the relative changes.

Starting in the IC-CDW phase at $T = 370\,\mathrm{K}$ [Fig.~\ref{ref:Figure4}(d)] the initial slope is higher for the \Eg spectrum than for \Alg symmetry. While the CDW gap started to open already at 554\,K around the $M$ points \cite{Sohrt2014}, which are highlighted by the \Eg vertex, the Fermi surface projected by the \Eg vertex continues to exist. Thus, we may interpret the different slopes as a manifestation of a momentum dependent gap in the IC-CDW phase and assume overall intensity effects to be symmetry-independent for all temperatures.  At $T = 352\,\mathrm{K}$ [Fig.~\ref{ref:Figure4}(e)] the slope for \Eg symmetry is substantially reduced to below the \Alg slope due to a strong increase of the CDW gap in the commensurate regions \cite{Sohrt2014} which emerge upon entering the NC-CDW phase. Further cooling also decreases the slope for the \Alg spectrum, as the Mott gap around the $\Gamma$ point starts to open within the continuously  growing C-CDW domains\cite{PhysRevB.56.13757,PhysRevB.94.201108}. Below $T = 270\,\mathrm{K}$ the initial slopes are identical for both symmetries and decrease with temperature. Apparently, the Mott gap opens up on the entire Fermi surface in direct correspondence with the increase of the resistivity by approximately an order of magnitude \cite{Wilson1975_AP24_117}.  Finally, at the lowest temperature close to $4\,\mathrm{K}$ the initial slopes drop to almost zero [Fig.~\ref{ref:Figure4}(g)] indicating vanishing conductivity or fully-gapped bands in the entire BZ.

Concomitantly, and actually more intuitive for the opening of a gap, we observe the loss of intensity in the Raman spectra below a threshold at an energy $\Omega_{\rm gap}$. {Below 30\,\wn the intensity is smaller than 0.2\,counts(mW\,s)$^{-1}$ [Fig.~\ref{ref:Figure4}(g)] and still smaller than 0.3\,counts(mW\,s)$^{-1}$ up to 1500\,\wn [Fig.~\ref{ref:Figure4}(h)].} For a superconductor or a CDW system the threshold is given by $2\Delta$, where $\Delta$ is the single-particle gap, and a pile-up of intensity for higher energies, $\Omega>2\Delta$ \cite{Devereaux:2007}. A pile-up of intensity cannot be observed here. Rather, the overall intensity is further reduced with decreasing temperature as shown in the Appendix in Figs.~\ref{ref:FigA1} and \ref{ref:FigA2}. In particular, the reduction occurs in distinct steps between the phases and continuous inside the phases with strongest effect in the C-CDW phase below approximately 210\,K [Fig.~\ref{ref:FigA1}]. In a system as clean as $1T$-TaS$_2$ the missing pile-up in the C-CDW phase is surprising and argues for an alternative interpretation.

In a Mott system, the gap persists to be observable but the pile-up is not a coherence phenomenon and has not been observed yet. In fact, the physics is quite different, and the conduction band is split symmetrically about the Fermi energy $E_{\rm F}$ into a lower and a upper Hubbard band. Thus in the case of Mott-Hubbard physics the experimental signatures are more like those expected for an insulator or semiconductor having a small gap, where at $T=0$ there is a range without intensity and an interband onset with a band-dependent shape. At finite temperature there are thermal excitations inside the gap. For $1T$-TaS$_2$ at the lowest accessible temperature, both symmetries exhibit a flat, nearly vanishing electronic continuum below a slightly symmetry-dependent threshold (superposed by the phonon lines at low energies).  Above the threshold a weakly structured increase is observed. We interpret this onset as the distance of the lower Hubbard band from the Fermi energy $E_{\rm F}$ or half of the distance between the lower and the upper Hubbard band, shown as vertical dashed lines at $1350-1550\,\wn \equiv 170-190\,\mathrm{meV}$ [Fig.~\ref{ref:Figure4}(h)]. The energy is in good agreement with gap obtained from the in-plane ARPES\cite{Sohrt2014}, scanning tunneling spectroscopy \cite{Skolimowski:2019} and  infrared spectroscopy\cite{Gasparov:2002}
which may be compared directly with our Raman results measured with in-plane polarizations. Upon increasing the temperature the size of the gap shrinks uniformly in both symmetries [Fig.~\ref{ref:Figure4}(i)] and may point to an onset above the C-CDW phase transition, consistent with the result indicated by the initial slope. However, we cannot track the development of the gap into the NC-CDW phase as an increasing contribution of luminescence (see Appendix~\ref{sec:luminescence}) overlaps with the Raman data. 

Recently, it was proposed on the basis of DFT calculations that $1T$-TaS$_2$ orders also along the $c$-axis perpendicular to the planes in the C-CDW state \cite{Darancet:2014,PhysRevLett.122.106404}. This quasi-1D coupling is unexpectedly strong and the resulting metallic band is predicted to have a width of approximately 0.5\,eV. For specific relative ordering of the ``star of David'' patterns along the $c$-axis this band develops a gap of 0.15\,eV at $E_{\rm F}$\cite{LeeSH:2019} which is intriguingly close to the various experimental observations. However, since our light polarizations are strictly in-plane, we have to conclude that the gap observed here (and presumably in the other experiments) is an in-plane gap. Our experiment can not detect out-of-plane gap. Thus, neither a quasi-metallic dispersion along the $c$-axis nor a gap in this band along $k_z$ may be excluded in the C-CDW phase. However, there is compelling evidence for a Mott-like gap in the layers rather than a CDW gap.

\section{Conclusions}

We have presented a study of the various charge-density-wave regimes in $1T$-TaS$_2$ by inelastic light scattering, supported by \textit{ab initio} calculations. The spectra of lattice excitations in the commensurate CDW (C-CDW) phase determine the unit cell symmetry to be $P3$, indicating trigonal or hexagonal stacking of the ``star-of-David'' structure. The high-temperature spectra of the incommensurate CDW (IC-CDW) state are dominated by a projection of the phonon density of states caused by either a significant electron-phonon coupling or, more likely, the superstructure. The intermediate nearly-commensurate (NC-CDW) phase is confirmed to be a mixed regime of commensurate and incommensurate regions contributing to the phonon spectra below an onset temperature $T_\mathrm{NC} \approx 352-360\,\mathrm{K}$, in good agreement with previously reported values. At the lowest measured temperatures, the observation of a virtually clean gap without a redistribution of spectral weight from low to high energies below $T_{\rm C}$ argues for the existence of a Mott metal-insulator transition at a temperature of order 100\,K. The magnitude of the gap is found to be $\Omega_{\rm gap} \approx 170-190\,\mathrm{meV}$ and has little symmetry, thus momentum, dependence in agreement with earlier ARPES results \cite{Rossnagel_2011}. At 200\,K, on the high-temperature end of the C-CDW phase, the gap shrinks to $\sim 60\%$ of its low-temperature value. Additionally, the progressive filling of the CDW gaps by thermal excitations is tracked via the initial slope of the spectra, and indicates that the Mott gap opens primarily on the parts of the Fermi surface closest to the $\Gamma$ point.

Our results demonstrate the potential of using inelastic light scattering to probe the momentum-dependence and energy-scale of changes in the electronic structure driven by low-temperature collective quantum phenomena. This opens perspectives to investigate the effect of hybridization on collective quantum phenomena in heterostructures composed of different 2D materials, e.g., alternating \textit{T} and \textit{H} monolayers as in the \textit{4Hb}-TaS$_2$ phase.\cite{Ribak2020}


\section*{Acknowledgements}
The authors acknowledge funding provided by the Institute of Physics Belgrade through the grant by the Ministry of Education, Science and Technological Development of the Republic of Serbia. The work was supported by the Science Fund of the Republic of Serbia, PROMIS, No. 6062656, StrainedFeSC, and by Research Foundation-Flanders (FWO). J.B. acknowledges support of a postdoctoral fellowship of the FWO, and of the Erasmus+ program for staff mobility and training (KA107, 2018) for a research stay at the Institute of Physics Belgrade, during which part of the work was carried out. The computational resources and services used for the first-principles calculations in this work were provided by the VSC (Flemish Supercomputer Center), funded by the FWO and the Flemish Government -- department EWI. Work at Brookhaven is supported by the U.S. DOE under Contract No. DESC0012704. A.~B. and R.~H. acknowledge support by the German research foundation (DFG) via projects Ha2071/12-1 and 107745057 – TRR 80 and by the DAAD via the project-related personal exchange program PPP with Serbia grant-no. 57449106.


%


\clearpage
\appendix

\begin{table}[htbp]
\small
\caption{\Alg and \Eg Raman mode energies experimentally obtained at \textit{T}= 4 K.}
\label{tableA1}
 \begin{tabular*}{0.48\textwidth}{@{\extracolsep{\fill}}lll}
\hline
$n_o$& $\omega_{\Alg} [\wn]$ & $\omega_{\Eg} [\wn]$  \\[1mm]\hline \\[-2mm]
1& 62.6 & 56.5  \\[1mm]
2& 73.3 & 63.3 \\[1mm]
3& 83.4 & 75.3  \\[1mm]
4& 114.9 & 82.0  \\[1mm]
5& 121.9 & 90.5  \\[1mm]
6& 129.5 & 101.1  \\[1mm]
7& 228.7 & 134.8 \\[1mm]
8& 244.1 & 244.0  \\[1mm]
9& 271.9 & 248.9  \\[1mm]
10& 284.2 & 257.5  \\[1mm]
11& 298.6 & 266.6  \\[1mm]
12& 307.2 & 278.3  \\[1mm]
13& 308.2 & 285.0 \\[1mm]
14& 313.0 & 292.9  \\[1mm]
15& 321.2 & 300.5 \\[1mm]
16& 324.2 & 332.7 \\[1mm]
17& 332.0 & 369.2  \\[1mm]
18& 367.2 & 392.6  \\[1mm]
19& 388.4 & 397.7  \\[1mm]
\hline
\end{tabular*}
\end{table}
\section{Raw data}
\label{sec:highenergyraw}

\begin{figure}[htbp]
 \centering
  \includegraphics[width=85mm]{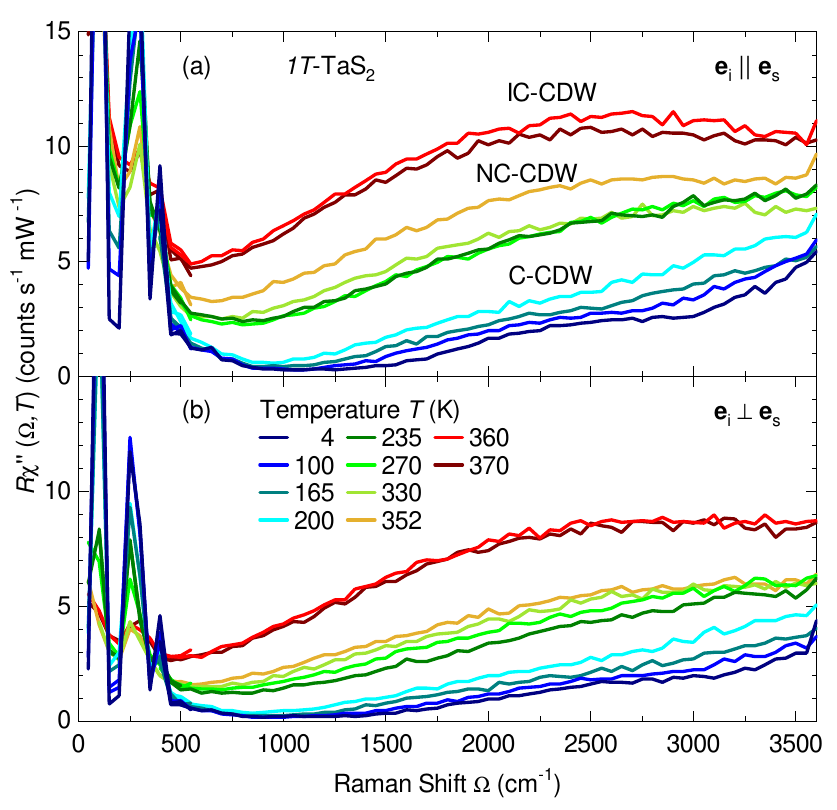}
  \caption{Raman spectra up to high energies for (a) parallel and (b) crossed polarizations of the incident and scattered light at temperatures as given in the legend.}
 \label{ref:FigA1}
\end{figure}

Figure~\ref{ref:FigA1} shows Raman spectra at temperatures ranging from $T = 4\,\mathrm{K}$ to $370\,\mathrm{K}$ for parallel [panel (a)] and crossed [panel (b)] in-plane light polarizations. The spectra were measured in steps of $\Delta\Omega = 50\,\wn$ and a resolution of $\sigma \approx 20\,\wn$. Therefore neither the shapes nor the positions of the phonon lines below 500\,\wn may be resolved. All spectra reach a minimum in the range from 500 to 1600\,wn. At energies above 500\,\wn the overall intensities are strongly temperature dependent and decreasing with decreasing temperature. Three clusters of spectra are well separated according to the phases they belong to.

In the C-CDW phase ($T \leq 200\,K$, blue lines) the spectra start to develop substructures at 1500 and 3000\,\wn. The spectra at 200\,K increase almost linearly with energy.
The spectra of the NC- and IC-CDW phases exhibit a broad maximum centered in the region of 2200-3200\,\wn which may be attributed to luminescence (see Appendix~\ref{sec:luminescence}). For clarification we measured a few spectra with various laser lines for excitation.

\section{Luminescence}
\label{sec:luminescence}

\begin{figure}[htbp]
 \centering
  \includegraphics[width=85mm]{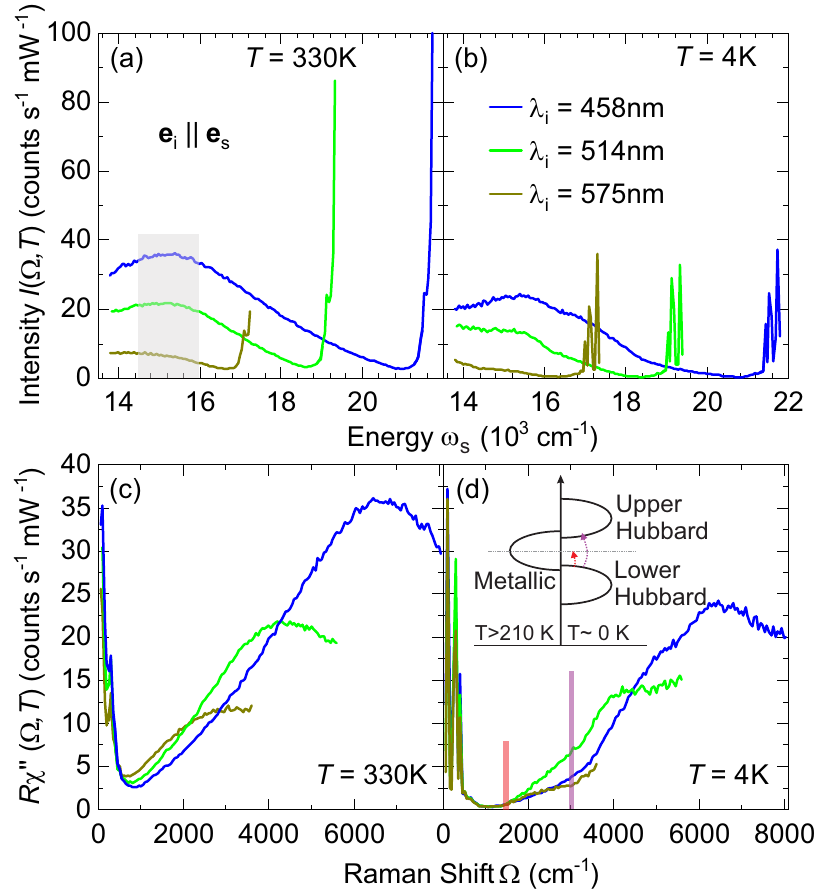}
  \caption{Luminescence contribution to the Raman data. (a-b) Intensity as a function of the absolute frequency for (a) $T=330\,\mathrm{K}$ and (b) $T=4\,\mathrm{K}$. The approximate peak maximum of the contribution attributed to luminescence is highlighted by the gray shaded area. (c-d) Raman susceptibility calculated from panels (a) and (b), respectively, shown as a function of frequency (Raman) shift. The luminescence peak appears at different Raman shifts depending on the wavelength of the laser light. At $T=4\,\mathrm{K}$ the spectra are identical up to 1600\,\wn for all laser light wavelengths.}
 \label{ref:FigA2}
\end{figure}

Figure~\ref{ref:FigA2} shows Raman spectra measured with parallel light polarizations for three different wavelengths $\lambda_\mathrm{i}$ of the incident laser light. Panels (a-b) depict the measured intensity $I$ (without the Bose factor) as a function of the absolute frequency $\tilde{\nu}$ of the scattered light.

At high temperature [$T = 330\,\mathrm{K}$, panel(a)] a broad peak  can be seen for all $\lambda_\mathrm{i}$ which is centered at a fixed frequency of 15200\,\wn of the scattered photons (grey shaded area). The peak intensity decreases for increasing $\lambda_\mathrm{i}$ (decreasing energy). Correspondingly, this peak's center depends on the laser wavelength in the spectra shown as a function of the Raman shift [panel (c)]. This behaviour indicates that the origin of this excitation is likely to be luminescence where transitions at fixed absolute final frequencies are expected.
\\
At low temperature [Fig.~\ref{ref:FigA2}(b)] we cannot find a structure at a fixed absolute energy any further. Rather, as already indicated in the main part, the spectra develop additional, yet weak, structures which are observable in all spectra but are particularly pronounced for blue excitation. For green and yellow excitation the spectral range of the spectrometer, limited to 732\,nm, is not wide enough for deeper insight into luminescence contributions (at energies different from those at high temperature) and no maximum common to all three spectra is observed. If these spectra are plotted as a function of the Raman shift the changes in slope at 1500 and 3000\,\wn are found to be in the same position for all $\lambda_\mathrm{i}$ values thus arguing for inelastic scattering rather than luminescence. Since we do currently not have the appropriate experimental tools for an in-depth study our interpretation is preliminary although supported by the observations in Fig.~\ref{ref:FigA2}(d).

As shown in the inset of Fig.~\ref{ref:FigA2}(d) we propose a scenario on the basis of Mott physics. In the C-CDW phase the reduced band width is not the largest energy any further and the Coulomb repulsion $U$ becomes relevant \cite{Fazekas_1979_229} and splits the conduction band into a lower and upper Hubbard band. We assume that the onset of scattering at 1500\,\wn corresponds to the distance of the highest energy of the lower Hubbard band to the Fermi energy $E_{\rm F}$. The second onset corresponds then to the distance between the highest energy of the lower Hubbard band and the lowest energy of the upper Hubbard band. An important question needs to be answered: Into which unoccupied states right above $E_{\rm F}$ the first process scatters electrons. We may speculate that some DOS is provided by the metallic band dispersing along $k_z$ or by the metallic domain walls between the different types of ordering patterns along the $c$-axis observed recently by tunneling spectroscopy \cite{Skolimowski:2019}. These quasi-1D domain walls would provide the states required for the onset of scattering at high energy but are topologically too small for providing enough density of states for measurable intensity at low energy [Fig. \ref{ref:Figure4}(g)] in a location-integrated experiment like Raman scattering.

\end{document}